\documentstyle[psfig]{aipproc2}
\begin{document}
\begin{tabular}{l}
\hbox to\hsize{ \hfill BROWN-HET-1165}\\
\hbox to \hsize{ \hfill TA-566}\\
\hbox to \hsize{ \hfill KIAS-P99003}\\ \end{tabular} 
\bigskip

\title
{Higgs Bounds in Three and Four Generation Scenarios}
\author{David Dooling$^{a}$, Kyungsik Kang$^{a}$ and Sin Kyu Kang$^{b}$}

\address{a. Department of Physics\\
Brown University, Providence RI 02912, USA \\
b. School of Physics \\
Korea Institute for Advanced Study \\
Seoul 130-012, Korea}
\maketitle
\begin{center}(Presented by \textbf{Kyungsik Kang}\footnote{Talk presented at the ``Physics at Run II'' Workshop on Supersymmetry/Higgs
Summary Meeting: 19-21 Nov, 1998
Fermi National Accelerator Laboratory, Batavia, Illinois})
\end{center}

\tightenlines
\begin{abstract}
In light of recent experimental results, we present updated bounds on the lightest Higgs boson mass in the Standard Model (SM) and in the Minimal Supersymmetric extension of the Standard Model (MSSM). The vacuum stability lower bound on the SM Higgs boson mass lies above the MSSM lightest Higgs boson mass upper bound for a large amount of SUSY parameter space. We postulate a fourth generation of fermions to see how the area of inconsistency changes, and discover that the MSSM is salvagable if a fourth generation is added to the MSSM (MSSM4).
\end{abstract}


\bigskip
\centerline {\rm\bf I. Introduction}
 The search for the Higgs boson being one of the major tasks along with that for supersymmetric sparticle and fourth generation fermions at future accelerators such as LEP200 and LHC makes it a theoretical priority to examine the bounds on the Higgs boson mass in the SM and its supersymmetric extension and to look for any distinctive features.
 The actual measurement of the Higgs boson mass could serve to exclude or at least to distinguish between the SM(3,4) and the MSSM(3,4) models for electroweak symmetry breaking.
Recently, bounds on the lightest Higgs boson mass were calculated in \cite{1,2,3,4,5,6,7,8,9}.
It was found that for a measured $M_{H}$ lying in a certain mass range, both the SM vacuum stability lower bound and the MSSM upper bound are violated, thus shaking our confidence in these theories just as the final member of the mass spectrum is observed.
One method of curing this apparent illness is to take a leap of faith by adding another fermion generation, to fortify these theories with another representation of the gauge group.
This additional matter content, for certain ranges of its mass values, has the desired effect of raising the MSSM3 upper bound above that of the SM lower bound and avoids the necessity of being forced to introduce completely new physics.

In this work, we use the latest LEP ElectroWeak Working Group data as well as the most recent experimental lower limits on the masses of fourth generation fermions to see how the Higgs boson mass bounds are affected.
With only the standard three generations, a violation of both the SM and the MSSM bounds occurs for a broard range of $M_{H}$ values, signalling the need for an additional generation.
Our presentation is organized as follows.
We first present a summary of our one-loop effective potential (EP) improved by the two-loop renormalization group equations (RGE) analysis.
An expression for the Higgs boson mass is derived up to the next-to-leading logarithm order.
Bounds on $M_{H}$ are obtained by imposing different boundary conditions on the Higgs self-coupling $\lambda$.
Finally, we present our results and also obtain information on the possible mass range of the fourth generation leptons, $M_{L}$.

\section{Effective Potential Approach }

As shown in \cite{10}, in order to calculate the Higgs boson mass 
 up to the next-to-leading logarithm approximation,
we must consider the one-loop EP 
improved by two-loop RGE for the $\beta$ and $\gamma$ functions of the
running coupling constants, masses and the $\phi$ field for the Higgs boson \cite{11}.

The two-loop RGE improved one loop EP of the SM4 is given by
\begin{equation}
 V_{1} = V_{(0)} + V_{(1)} 
\end{equation}
where
\begin{eqnarray}
 V_{(0)} &=& -\frac{1}{2}m^{2}(t)\phi_{c}^{2}(t) + \frac{1}{24}\lambda(t)
              \phi_{c}^{4}(t) \\
V_{(1)} &=& \sum_{i=1}^{5} \left(-\frac{\kappa_{i}}{64\pi^{2}}\right)
            h_{i}^4(t)\phi_{c}^{4}(t)
            \left[\ln\frac{h_{i}^{2}(t)\zeta^{2}(t)}{2}-\frac{3}{2}\right] 
\end{eqnarray}
Here $\phi_c$ is the classical field corresponding to the physical
Higgs boson $\phi$, $i = (t,T,B,N,E)$
, $~\kappa_{i}=3$ for $i = (t,T,B)$ and $\kappa_{i}=1$ for $i = (N,E)$ and $h_{i}$ is the Yukawa coupling of the i$^{th}$ fermion to the Higgs field.
In addition,
\begin{eqnarray}
\zeta(t) &=& \left(-\int_{0}^{t} \gamma_{\phi}(t^{\prime})dt^{\prime}\right)
\\ 
\phi_{c}(t) &=& \phi_{c} \zeta(t)
\end{eqnarray}
and
\begin{equation}
\mu(t)=\mu\exp(t)
\end{equation}
where $\mu $ is a fixed scale.

Starting with the above expression for the SM(3,4) EP, one may follow the analysis in \cite{1,4} and obtain the expression for the Higgs boson mass:

\begin{eqnarray}
m_{\phi}^{2} &=& \frac{1}{3}\lambda v^{2}+\frac{\hbar v^{2}}{16 \pi^{2}}\left\{ \frac{\lambda^{2}}{3} +2\lambda (h_{t}^{2}+h_{T}^{2}+h_{B}^{2}) +\frac{2\lambda}{3}(h_{E}^{2}+h_{N}^{2})\right. \nonumber \\
& & -\frac{\lambda}{2}(3 g_{2}^{2}+g_{1}^{2}) + \frac{9}{8}g_{2}^{4} +\frac{3}{4}g_{1}^{2}g_{2}^{2} + \frac{3}{8}g_{1}^{4} \nonumber \\
& &-6h_{t}^{4}\ln\frac{h_{t}^{2}\zeta^{2}}{2} -6h_{T}^{4}\ln\frac{h_{T}^{2}\zeta^{2}}{2}-6h_{B}^{4}\ln\frac{h_{B}^{2}\zeta^{2}}{2} \nonumber \\
& & \left. -2h_{E}^{4}\ln\frac{h_{E}^{2}\zeta^{2}}{2}-2h_{N}^{4}\ln\frac{h_{N}^{2}\zeta^{2}}{2} \right\} + O(\hbar^{2})
\end{eqnarray}
where the three generation case is obtained by simply letting the fourth generation Yukawa couplings go to zero.

Following the method similar of Kodaira et al. \cite{4}, we arrive at the appropiate energy scale at which to evaluate $\frac{\partial^{2} V}{\partial \phi_{c}(t)^{2}}$ by requiring $\frac{\partial V}{\partial \phi_{c}(t_{v})} =$ 0 at the scale $t_{v}$ where $\phi_{c}(t_{v}) = v = (\sqrt{2} G_{F})^{-\frac{1}{2}}$ = 246 GeV.
$t_{v}$ is found to satisify:
\begin{displaymath}
t_{v} = \ln \frac{v}{\mu} + \int_{0}^{t_{v}} \gamma_{\phi}(t^{\prime})dt^{\prime}.
\end{displaymath}
We then evaluate the first and second derivatives of the EP at the scale $t_{v}$ where $\phi_{c}(t_{v}) = v$.
In the above relation satisfied by $t_{v}$, $\mu$ is a fixed, constant mass scale.
In the MSSM theories, we will take $\mu$ to be $M_{susy}=1$ or 10 TeV, while the SM lower bounds will be derived after choosing $\mu= \Lambda= 10^{19}$ GeV.

 We define the running Higgs mass as:
\begin{displaymath}
m_{h}^{2}(t)= \frac{m_{h}^{2}(t_{v}) \zeta^{2}(t_{v})}{\zeta^{2}(t)}.
\end{displaymath}
The physical, pole masses are related to the running masses via the following equations:
\begin{displaymath}
M_{i}= \left[ 1 +\beta_{i} \frac{4 \alpha_{3}(M_{i})}{3 \pi} \right] m_{i}(M_{i})
\end{displaymath}
\begin{displaymath}
M_{H}^{2} = m_{h}^{2}(t) + Re\Pi(M_{H}^{2}) - Re\Pi(0)
\end{displaymath}
where $\beta_{i}$ = 0 for i = (N,E) and = 1 for i = (t,T,B).
$\Pi(q^{2})$ is the renormalized electroweak self-energy of the Higgs boson.

The method of solving the RGE and the appropriate boundary conditions for the couplings is explained in \cite{1}.
In this update, we use $M_{Z} = 91.1867$ GeV and $\alpha_{3}(M_{Z}) = .119$.

\section{Bounds on $M_{H}$}

Let us first discuss the procedure for determining a lower bound on the Higgs boson mass in the SM \cite{5,12}
Working with the two-loop RGE requires the imposition of one-loop boundary conditions on the running parameters.
As pointed out by Casas et al. \cite{5,7}, the necessary condition for vacuum stability is derived from requiring that the effective coupling $\tilde{\lambda}(\mu)>$ 0  rather than $\lambda > 0$ for $\mu(t) < \Lambda$, where $\Lambda$ is the cut-off beyond which the SM is no longer valid. 
The effective coupling $\tilde{\lambda}$ in the SM4 is defined as:
\begin{displaymath}
\tilde{\lambda}=\frac{\lambda}{3} -\frac{1}{16 \pi^{2}}\left\{ \sum_{i=1}^{5} 2 \kappa_{i} h_{i}^{4} \left[ \ln \frac{h_{i}^{2}}{2} - 1 \right] \right\}
\end{displaymath}
where the three generation case is simply the same as the above expression without the fourth generation Yukawa coupling contributions.
Choosing $\Lambda = 10^{19}$ GeV and $M_{top} = 172$ GeV, we arrive at a vacuum stability lower bound on $M_{h}$ of $\sim$ 134 GeV for the SM with three generations.
Allowing $M_{top}$ to be as large as 179 GeV increases the lower bound on $M_{H}$ to $\sim$ 150 GeV.

To compute the MSSM upper bound on $M_{H}$, we assume that all of the sparticles have masses $O(M_{susy})$ or greater and that of the two Higgs isodoublets of the MSSM, one linear combination is massive, also with a mass of $O(M_{susy})$ or greater, while the other linear combination, orthogonal to the first, has a mass of the order of weak-scale symmetry breaking.
With these two assumptions, it is clear that below the supersymmetry breaking scale $M_{susy}$, the effective theory is the SM.
This fact enables us to use the SM effective potential for the Higgs boson when we treat the lightest Higgs boson in the MSSM.

In the MSSM(3,4), the boundary condition for $\lambda$ at $M_{susy}$ is
\begin{displaymath}
\frac{\lambda}{3}(M_{susy})=\frac{1}{4}\left[g_{1}^{2}(M_{susy})+g_{2}^{2}(M_{susy})\right]\cos^{2}(2\beta)+\frac{\kappa_{i}h_{i}^{4}(M_{susy})}{16 \pi^{2}}\left( 2\frac{X_{i}}{M_{susy}^{2}}-\frac{X_{i}^{4}}{6 M_{susy}^{4}} \right)
\end{displaymath}
where $\kappa_{i}$ = 3 for $i = (t,T,B)$ and $\kappa_{i}$ = 1 for $i = (N,E)$ and $X_{i}$ is the supersymmetric mixing parameter for the ith fermion.
Zero threshold corrections correspond to $X_{i}$ = 0.
Maximum threshold corrections occur for $X_{i} = 6 M_{susy}^{2}$.
\bigskip
\bigskip
\vglue -1cm
\hglue -2cm
\psfig{figure=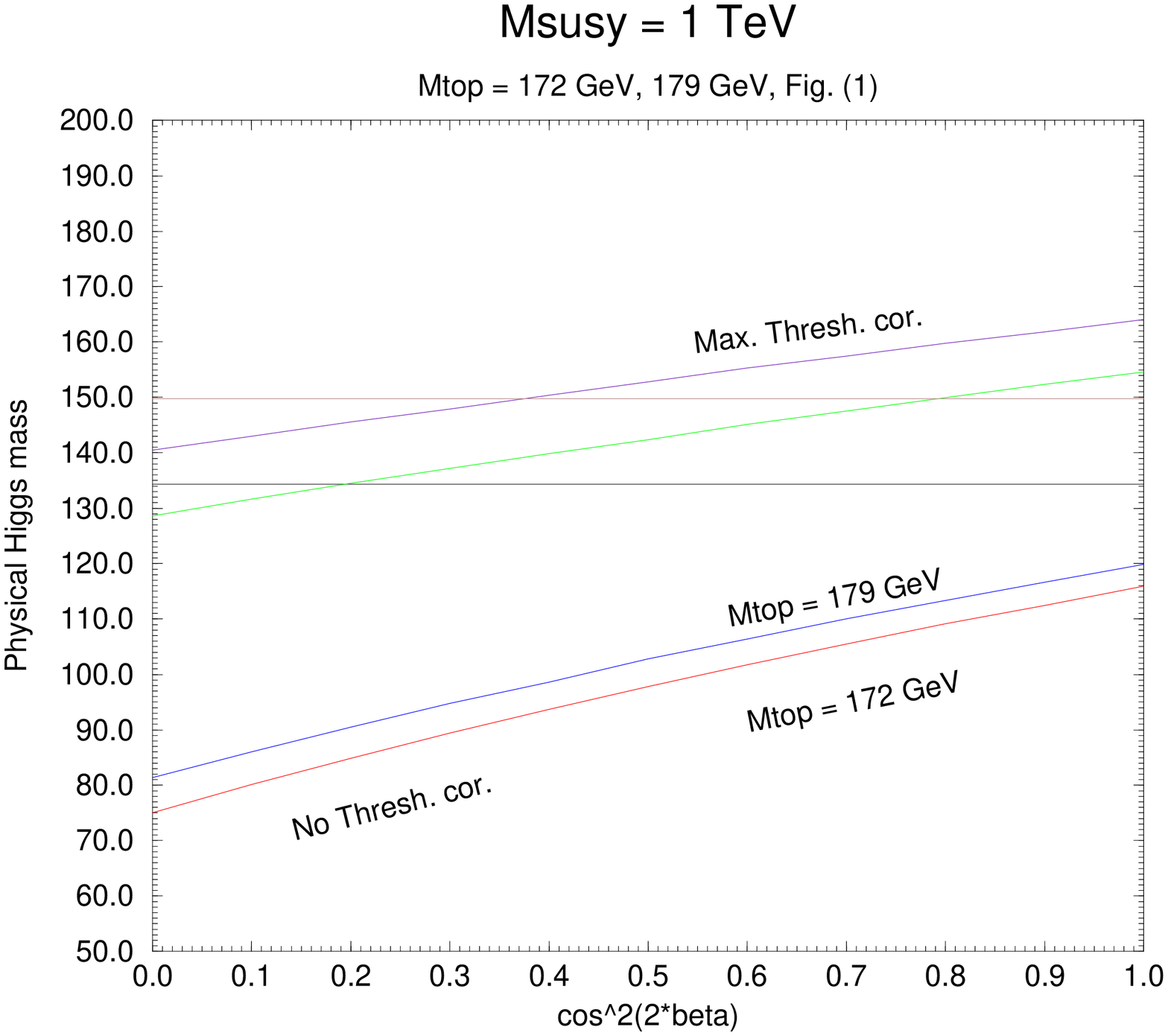,height=8cm,angle=-90}\hglue 1cm
\psfig{figure=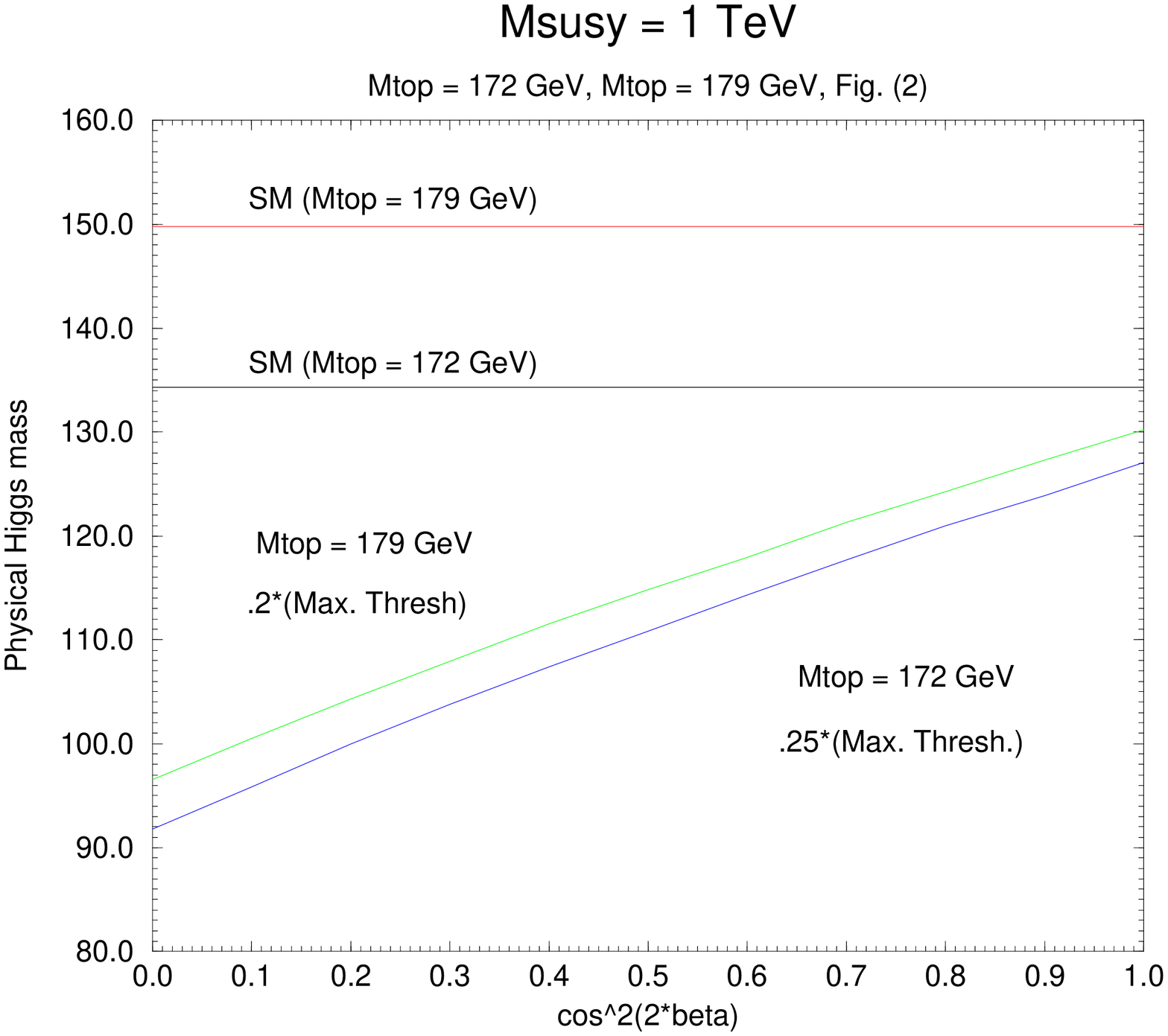,height=8cm,angle=-90}
\hglue 3.5cm (1)\hglue 10cm (2)\\
\begin{quote}
\scriptsize Figure 1: The lightest Higgs boson mass $M_{H}$ as a function of $\cos^{2}(2\beta)$.
The bottom two curves correspond to MSSM upper bounds with no threshold corrections, for $M_{top}$ = 172 GeV and 179 GeV, respectively.
 The two upper curves correspond to MSSM upper bounds with maximum threshold corrections, for $M_{top}$ = 172 GeV and 179 GeV, respectively.
The two horizontal lines are the $\cos^{2}(2\beta)$-independent SM3 vacuum stability bounds.
The lower horizontal line corresponds to $M_{top}$ = 172 GeV, while the other horizontal line was computed with $M_{top}$ = 179 GeV.
Figure 2: Same as Figure 1, but now the MSSM bounds correspond to the minimal threshold corrections consistent with the experimental lower limit on $M_{H}$
\end{quote}
\normalsize
In Fig. (1) we present our numerical two-loop results for the lightest Higgs boson mass bounds in the SM and the MSSM3 as a function of the supersymmetric parameter $\cos^{2}(2\beta)$.
The bottom two curves correspond to the MSSM3 upper bound for the two cases $M_{top} = 172$ GeV and the slightly greater upper bound that results when $M_{top}$ = 179 GeV and with no threshold corrections.
When the case of maximum threshold corrections is considered, these two curves are translated upwards by $\sim$ 55 GeV - 60 GeV, illustrating the strong dependence of the upper bound on the precise value of the threshold corrections.
Yet even with such a dramatic increase in the upper bounds with increasing threshold corrections, we observe that the SM lower bound exceeds the MSSM upper bound for $M_{top} = 172$ GeV and $ 0 < \cos^{2}(2\beta) < .2$ for all values of the threshold correction contribution.
Similarly, for $M_{top} = 179$ GeV, the troublesome situation is only exacerbated, as the SM lower bound exceeds the MSSM upper bound for $0 < \cos^{2}(2\beta) < .38$ independent of the threshold corrections.

In Fig.(2) we present the problem more clearly.
Taking into account the present experimental lower limit on $M_{H}$ of $\sim$ 90 GeV at 95$\%$ CL, we find the value of the threshold correction that gives a smallest upper bound consistent with the experimental lower limit.
Clearly, for this phenomenologically determined lower limit of the threshold contributions, there is a large area in $M_{H} \times \cos^{2}(2\beta)$ space that is inconsistent with both the SM and the MSSM.
For $M_{top} = 172$ GeV, the region 92 GeV $ < M_{H} < $ 134 GeV invalidates both theories independent of $\cos^{2}(2\beta)$, while for $M_{top} = 179$ GeV, the range of mutual invalidiation is 92 GeV $ < M_{H} < $ 150 GeV.

\section{Fourth Generation}

To resolve the above conundrum, one would like to either raise the MSSM upper bounds, lower the SM lower bounds, or both.
Adding more massive fermions to the theory only increases both bounds, so it is readily apparent that the way out of the area of inconsistency is to consider the MSSM4 and see if the additional matter of the MSSM4 results in MSSM4 upper bounds that exceed the SM3 lower bounds.

We now discuss restrictions on the possible fourth generation fermion masses \cite{2,13,14,15}.
The close agreement betweeen the direct measurements of the top quark at the Tevatron and its indirect determination from the global fits of precision electroweak data including radiative corrections within the framework of the SM imply that there is no significant violation of the isospin symmetry for the extra generation.
Thus the masses of the fourth generation isopartners must be very close to degenerate \cite{14}; i.e.
\begin{displaymath}
\frac{\|M_{T}^{2}-M_{B}^{2}\|}{M_{Z}^{2}} \lesssim 1, \frac{\|M_{E}^{2}-M_{N}^{2}\|}{M_{Z}^{2}} \lesssim 1
\end{displaymath}
Recently, the limit on the masses of the extra neutral and charged lepton masses, $M_{N}$ and $M_{E}$, has been improved by LEP1.5 to $M_{N} > 59$ GeV and $M_{E} > 62$ GeV.
Also, CDF has yielded a lower bound on $M_{B}$ of $\sim$ 140 GeV.

In our previous work, we considered a completely degenerate fourth generation of fermions with mass $m_{4}$.
We derived an upper bound on $m_{4}$ in the MSSM4 by demanding pertubative validity of all the couplings out to the GUT scale \cite{16}.
This constraint led to an upper bound on $m_{4}$ of $\sim$ 110 GeV.
The above experimental lower limit on $M_{B}$ naturally forces us to now a consider a fourth generation where degeneracy only holds among the isodoublets seperately.
We therefore consider a fourth generation with masses $M_{L}$ and $M_{Q}$.

In Fig.(3), we present the SM lower bound, the MSSM4 upper bound with the fourth generation masses at their experimental lower limits and with fourth generation masses large enough to remove the problem area for all values of $\cos^{2}(2\beta)$.
The MSSM bounds were calculated with no threshold corrections, and $M_{top}$ is fixed at 172 GeV.
Fig.(4) shows the same information for $M_{top}$ = 179 GeV.
The MSSM4 upper bounds are much more sensitive to $M_{Q}$ than they are to $M_{L}$.
This qualitative behaviour is readily understood from inspection of the equation for $m_{\phi}^{2}$.
For this reason, it is necessary to increase $M_{Q}$ appropriately in order to generate a MSSM4 upper bound that is greater than the SM lower bound for all values of $\cos^{2}(2\beta)$.
In fact, keeping $M_{Q}$ at 146 GeV and allowing $M_{L}$ to be 110 GeV does not resolve the problem.
But increasing both $M_{Q}$ and $M_{L}$ as indicated in the figures does remove the problem.
Because all of the bounds increase as $M_{L}$ and $M_{Q}$ increase, and because the upper bounds on $m_{4}$ from the previous work are saturated when the masses of the fourth generation reach some critical values from below, we can conclude that $M_{L}$ must still be $< 110$ GeV.
This conclusion follows because it is $h_{N}$ that violates pertubative validity, so in the non-degenerate case, it is $M_{L}$ that must still respect this upper bound if gauge coupling unification is still to be achieved in the MSSM4.

\hglue -2cm
\psfig{figure=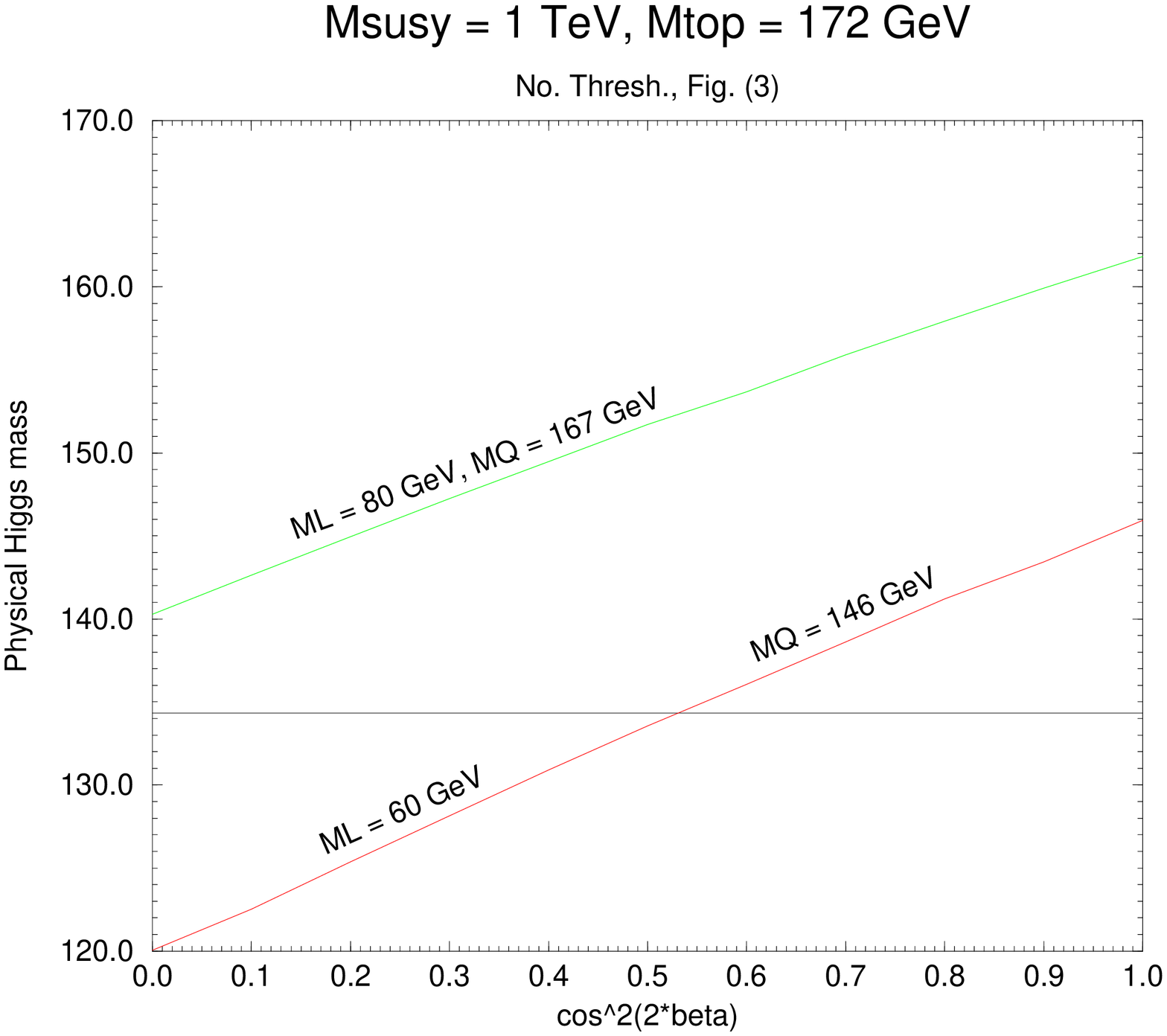,height=8cm,angle=-90}\hglue 1cm
\psfig{figure=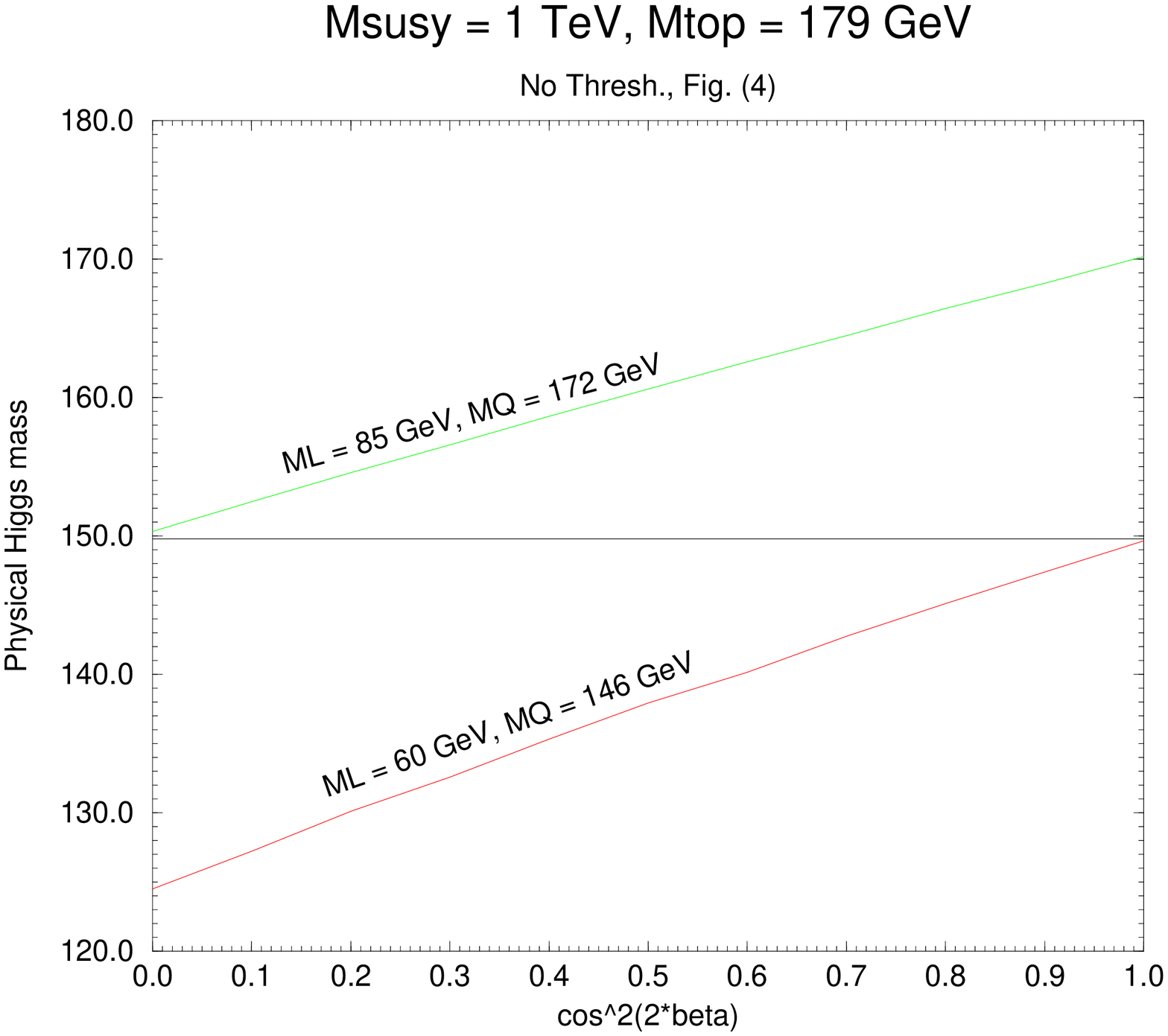,height=8cm,angle=-90}
\hglue 3.5cm (3)\hglue 10cm (4)\\
\begin{quote}
\scriptsize Figure 3: Plots of the physical Higgs boson mass as a function of $\cos^{2}(2\beta)$. The $\cos^{2}(2\beta)$-independent flat line is the MSSM3 vacuum stability lower bound for $M_{top}$ = 172 GeV. The lower curve is the MSSM4 upper bound for the same value of $M_{top}$, no threshold corrections and the indicated values for $M_{L}$ and $M_{Q}$. Similarly for the upper curve. Figure 4: Same as Figure 3, but with $M_{top}$ = 179 GeV.
\end{quote}
\normalsize
\section{CONCLUDING REMARKS}

In conclusion, we have studied the upper bounds on the lightest Higgs boson mass $M_{H}$ in the MSSM with four generations by solving the two-loop RGEs and using the one-loop EP.
We have considered a fourth generation of quarks and leptons with degenerate masses $M_{Q}$ and $M_{L}$.
For certain values of $M_{Q}$ and $M_{L}$, the area of mutual inconsistency between the SM and MSSM3 Higgs mass bounds is found to be consistent with the MSSM4 upper bounds.

\section{ACKNOWLEDGEMENTS}

We wish to thank M. Machacek and M. Vaughn for helpful discussions concerning the RGE used in this investigation.
Support for this work was provided in part by U.S. Dept. of Energy Contract DE-FG-02-91ER40688-Task A.

\end{document}